\documentclass[letterpaper]{article}

\usepackage[T1]{fontenc}

\usepackage{geometry}
\usepackage{setspace}
\usepackage{caption}

\usepackage[style = chem-acs]{biblatex}
\usepackage{graphicx}
\usepackage{float}
\newfloat{scheme}{htbp}{los}
\floatname{scheme}{Scheme}
\floatname{chart}{Chart}
\newfloat{graph}{htbp}{loh}

\usepackage{chemformula} % Formulas using \ch{}
\usepackage[version = 4]{mhchem} % Formulas using \ce{}

\usepackage{authblk}
\author[1,5]{Zihao Liu}
\author[2]{Yunzhe Ke}
\author[4,5]{Hajar Komekoz}
\author[2]{Shuai Gu}
\author[3,5]{Rico Ehrler}
\author[8]{Chen Wang}
\author[3,5]{Olav Hellwig}
\author[6,7]{Frank Ortmann}
\author[5]{Dietrich R. T. Zahn}
\author[8]{Carsten Deibel}
\author[4,5]{Michael Sommer}
\author[2]{Shuaishuai Ding*}
\author[1,5]{Georgeta Salvan*}

\affil[1]{Magnetooptics, Institute of Physics, Chemnitz University of Technology, Chemnitz, Germany}

\affil[2]{State Key Laboratory of Advanced Materials for Intelligent Sensing \& Tianjin Key Laboratory of Molecular Optoelectronic Sciences, Department of Chemistry, Institute of Molecular Aggregation Science, Tianjin University, Tianjin 300072, China}

\affil[3]{Magnetic Functional Materials, Institute of Physics, Chemnitz University of Technology, Chemnitz, Germany}

\affil[4]{Polymer Chemistry, Institute of Chemistry, Chemnitz University of Technology, Chemnitz, Germany}

\affil[5]{Research Center for Materials, Architectures and Integration of Nanomembranes (MAIN), Chemnitz University of Technology, Chemnitz, Germany}

\affil[6]{Department of Chemistry, TUM School of Natural Sciences, Technical University of Munich, Garching b. München, Germany}

\affil[7]{Atomistic Modeling Center, Munich Data Science Institute, Technical University of Munich, Garching b. München, Germany}

\affil[8]{Optics and Photonics of Condensed Matter, Institute of Physics, Chemnitz University of Technology, Chemnitz, Germany}

\title{Interfacial Effects of Thermal Reorganization on Charge and Spin Transport in Organic Spin Valves}
\date{*Email: salvan@physik.tu-chemnitz.de; dingshuaishuai@tju.edu.cn
}

\begin{document}

\maketitle

\begin{abstract}
  Organic semiconductors are attractive for spintronic applications owing to their weak spin--orbit coupling, small hyperfine interactions, mechanical flexibility, and tunable electronic properties. Organic spin valves (OSVs) provide a representative platform for investigating spin-dependent transport. A typical OSV consists of an organic semiconductor between two ferromagnetic electrodes, with spin transport evaluated by the magnetoresistance (MR) response. However, the MR response is highly sensitive to the ferromagnetic electrode/organic semiconductor interface (spinterface). Variations in interfacial morphology, chemical interaction, energy-level alignment, and spin polarization can strongly affect spin injection, transport, and detection, making reproducible OSV performance difficult to achieve. Here, we investigate how in situ thermal annealing affects the electrical and magnetotransport properties of complete NDI-based OSVs. Two top-electrode configurations are compared: directly evaporated Co/Au and transferred Gr(graphene)/Co/Au. Both as-cast devices exhibit clear MR responses, with a stronger absolute response observed for the transferred-electrode device. After in situ annealing, the resistance of both device types increases substantially, and no reproducible MR response is resolved, despite the enhanced structural order observed in companion films and an approximately 36-fold increase in the effective vertical conductivity extracted from electron-only devices. This contrasting behavior indicates that improved bulk order alone is insufficient to ensure enhanced spin-valve performance. Instead, annealing-induced changes in interfacial contact, contact homogeneity, and local current distribution can become dominant in determining the complete-device response. These results emphasize the critical role of spinterface stability and interface engineering in achieving reliable and thermally robust OSV operation.
\end{abstract}

\section*{Keywords}

organic spin valves; spinterfaces; thermal annealing; magnetoresistance; electrode transfer; naphthalene diimide oligomers

\section*{Abbreviations}

OSV, organic spin valve; MR, magnetoresistance; FM, ferromagnetic; NDI, naphthalene diimide; T2, bithiophene;EH, 2-ethylhexyl; also used herein as shorthand for the T2-(NDI-T2)$_2$-2EH oligomer; Gr, graphene; AFM, atomic force microscopy; XRR, X-ray reflectivity; SLD, scattering length density; RMS, root-mean-square; NMR, nuclear magnetic resonance; EOD, electron-only device; ITO, indium tin oxide; ZnO, zinc oxide; PDINO, perylene diimide functionalized with amino N-oxide; PPMS, Physical Property Measurement System; SI, Supporting Information.

\section{Introduction}

Over the past few decades, spintronics has provided new opportunities for information storage, transmission, and processing by employing both the charge and spin of electrons~\cite{Naber2007,Dediu2009,Sun2014}. Among various spintronic devices, the spin valve is a fundamental device structure for studying spin-dependent transport. It typically consists of two ferromagnetic (FM) electrodes and a nonmagnetic transport layer sandwiched between them, where the device resistance depends on the relative orientation of the two ferromagnetic electrode magnetizations~\cite{Naber2007,Xiong2004}. Organic semiconductors typically exhibit weak spin--orbit coupling and small hyperfine interactions, which help extend spin lifetime, making them promising spin-transport materials~\cite{Naber2007,Dediu2009,Watanabe2014,Nguyen2010,Nuccio2013SpinOrbit}. Organic spin valves (OSVs) further convert these material properties into directly measurable device-level magnetoresistance (MR) responses and provide an effective platform for studying the relationships among molecular stacking and spin-dependent transport~\cite{Drew2009SpinDiffusion}. Recent studies have therefore increasingly explored molecular design to regulate spin relaxation and interface engineering to control spin injection and device functionality~\cite{Qin2025SpinLifetime,Liu2025OrganicSpintronics,Zhang2025MultilevelOSV}. However, the intrinsic advantages of the materials do not necessarily translate into stable and reproducible device performance. Achieving reliable spin-dependent transport in OSVs remains a challenge~\cite{Sun2014,Li2021Innovation}. This issue arises largely from the ferromagnet/organic semiconductor interface (spinterface), because spin injection and spin detection processes are highly sensitive to the structure and electronic properties of this interface~\cite{Barraud2010Unravelling,Cinchetti2017Activating,Raman2013Interface}. Even minor changes in factors such as interface topography, roughness, chemical bonding, and metal penetration can alter the interface spin polarization and carrier injection conditions, thereby significantly affecting the ultimately measured MR response~\cite{Barraud2010Unravelling,Cinchetti2017Activating,Raman2013Interface,Guo2024Strain,Ding2018Reliable,Ciudad2014Interface}.

Specifically, the spinterface is affected not only by the fabrication process of the top ferromagnetic electrode but also by the surface morphology of the organic semiconductor~\cite{Guo2024Strain,Ding2018Reliable,Luo2023MolecularPacking,Droghetti2016Dynamic}. During direct metal evaporation, metal atoms may penetrate the organic layer and induce localized structure damage~\cite{Ding2018Reliable,Zhan2008Energy}. Therefore, top-electrode transfer techniques have been employed to minimize such fabrication-induced damage, thereby improving device reliability and reproducibility~\cite{Guo2024Strain,Ding2018Reliable}. Related van der Waals top-contact strategies in organic electronic devices
have likewise demonstrated that preformed electrodes can preserve the
underlying semiconductor surface while avoiding deposition-induced
damage~\cite{Zhang2023VdWContact}. When external damage caused by electrode deposition is reduced, the contact quality at the spinterface will become more dependent on the intrinsic surface characteristics of the organic semiconductor. Surface roughness, local variations in molecular packing, and non-uniform topography distribution may lead to spatially inhomogeneous contact conditions at the spinterface~\cite{Ye2025SmartMat}. These local contact variations may introduce additional spin scattering and nonuniform current pathways, thereby reducing the efficiency of spin injection and detection~\cite{Guo2024Strain,Ding2018Reliable,Luo2023MolecularPacking,Droghetti2016Dynamic,Zhang2015Rubrene}. At the molecular scale, changes in adsorption geometry and orbital hybridization at the spinterface can further modify the local
spin polarization, emphasizing the strong sensitivity of spin injection to the
microscopic interfacial structure~\cite{Atodiresei2010SpinPolarization,
Djeghloul2013Spinterface}. Previous studies have mainly focused on optimizing the spinterface by controlling interfacial oxidation or tuning energy-level alignment
~\cite{Zhan2008Energy,Bergenti2018Oxygen,Riminucci2019Oxygen,
Galbiati2014Spinterface, Zhang2023SmallSpinterface}. Such strategies typically rely on pre-designed interface structures and evaluate the influence of the interface on spin transport by comparing performance among different devices~\cite{Riminucci2019Oxygen,Bergenti2018Oxygen,Bergenti2019Spinterface}. Therefore, it remains necessary to use the same organic materials and thermal processing steps to compare the direct evaporation method with the electrode transfer method. Doing this enables a systematic study, within complete devices, of how variations in local interface contact conditions affect spin injection, spin transport, and the resulting MR response.

Based on these considerations, we selected the discrete n-type conjugated oligomer T2-(NDI-T2)$_2$-2EH, which consists of alternating electron-accepting naphthalene diimide (NDI) and electron-donating bithiophene (T2) units, as the active material~\cite{Matsidik2023Evolution}. In the molecular name, EH denotes the 2-ethylhexyl side chains; for brevity, the complete oligomer is hereafter referred to as EH. As demonstrated in our recent work, this structurally well-defined oligomer exhibits n-type electron transport and undergoes pronounced molecular reorganization upon thermal annealing~\cite{Liu2026SideChain}, consistent with structural changes reported for related NDI--T2 oligomers~\cite{Ehm2025Alignment}. It is therefore well suited for investigating thermally induced structural evolution and its effects on charge and spin transport. More generally, modifications in molecular arrangement can simultaneously influence charge-carrier transport and spin-dependent transport in conjugated semiconductors~\cite{Li2019StructureTransport}. Annealing provides a means of modifying the molecular organization of the organic layer without changing its chemical composition. The associated structural evolution was characterized using independently prepared, electrode-free companion films subjected to the same annealing temperature and duration as the OSVs. Electrical and magnetotransport measurements were performed on the same completed OSV before and after annealing. Two EH-based device architectures fabricated using different methods were compared: a NiFe/EH/Co/Au OSV formed by direct top-electrode evaporation and a NiFe/EH/Gr/Co/Au OSV formed using the transferred Gr/Co/Au top electrode~\cite{Ding2018Reliable,Ke2026Temperature}. This comparative design allows us to examine the intrinsic structural response of the material. It also enables comparison with the annealing-induced changes in electrical and magnetotransport behavior of the two devices with different fabrication processes. The results show that annealing enhances the structural order of the organic layer and increases its effective vertical conductivity by approximately 36-fold, whereas the resistances of both completed OSVs increase and their reproducible MR responses become absent. Together with the increased roughness observed in the companion films, this contrasting behavior is consistent with annealing-induced modifications of interfacial contact and local current distribution. These findings suggest that improved bulk molecular order does not necessarily lead to enhanced spin-dependent transport; in OSVs, spinterface quality, particularly contact continuity and homogeneity, can play a more critical role in determining the magnetotransport response.

\section{Results and discussion}

\begin{figure}[h]
      \centering
      \includegraphics[width=\linewidth]{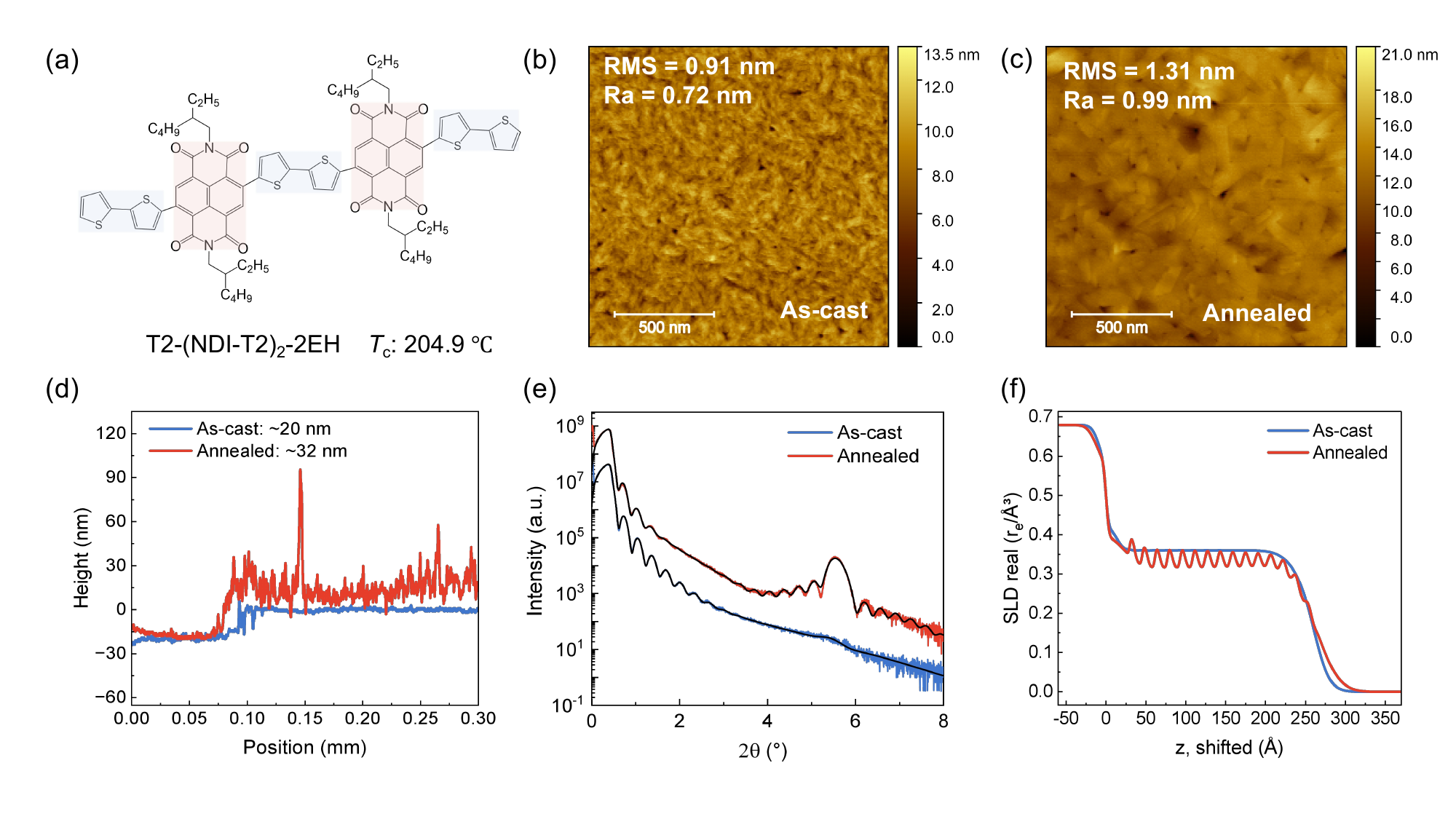}
      \caption{
  Structure and morphology of T2--(NDI--T2)$_2$--2EH films in the as-cast and annealed states. (a) Molecular structure. (b,c) AFM images of the same sample in the (b) as-cast and (c) annealed states. (d) Step-height profiles of the same film in the as-cast and annealed states. The annealed state was obtained by in situ thermal annealing. (e) XRR curves with fits. (f) Corresponding SLD profiles.}
      \label{structureandmorphology}
\end{figure}

The molecular structure and thermal transition temperature ($T_{\mathrm{c}} = 204.9~^{\circ}\mathrm{C}$) of T2-(NDI-T2)$_2$-2EH are shown in Figure~\ref{structureandmorphology}a~\cite{Matsidik2023Evolution}. To examine annealing-induced morphological changes directly, the same film specimen was characterized before and after in situ annealing at $208~^{\circ}\mathrm{C}$ for 10~min. As revealed by the atomic force microscopy (AFM) images, the as-cast film exhibited a relatively smooth and uniform surface without clearly resolved domains, with root mean square (RMS) roughness and profile roughness parameter $R_{\mathrm{a}}$ showing values of 0.91 and 0.72~nm, respectively. After annealing, the surface morphology reorganized into distinct elongated domains, accompanied by an increase in the RMS and $R_{\mathrm{a}}$ roughness to 1.31 and 0.99~nm, respectively (Figure~\ref{structureandmorphology}b,c). The step-height profiles measured on the same film before and after annealing (Figure~\ref{structureandmorphology}d) show apparent heights of approximately 20 and 32 nm, respectively. This change may be associated with thermally induced morphological reorganization, although local step-height measurements alone do not establish a uniform thickness increase across the entire film. Such behavior is consistent with the general ability of thermal treatment to reorganize molecular packing and crystallite orientation in conjugated semiconductor films~\cite{Seifrid2018Orientation,Son2021Orientation}. The larger height fluctuations across the annealed film are consistent with the formation of surface domains and the increased RMS roughness observed in Figure~\ref{structureandmorphology}c. While AFM reflects the evolution of the surface morphology, changes in the
internal organization of the film were further evaluated by X-ray
reflectivity (XRR). The as-cast film exhibited thickness fringes without a pronounced higher-angle reflection, whereas the annealed film showed an additional reflection at approximately $5.5^{\circ}$, together with more distinct structural modulations between $4^{\circ}$ and $6^{\circ}$ (Figure~\ref{structureandmorphology}e). Consistently, the scattering length density (SLD) profile changed from a relatively uniform distribution in the as-cast film to a periodic modulation throughout the annealed organic layer (Figure~\ref{structureandmorphology}f). Taken together, these results indicate that annealing increases the surface roughness of the film while enhancing its structural order, which also provides a structural reference for evaluating the electrical and magnetotransport responses of completed OSVs subjected to the same thermal treatment.

\begin{figure}[h]
      \centering
      \includegraphics[width=\linewidth]{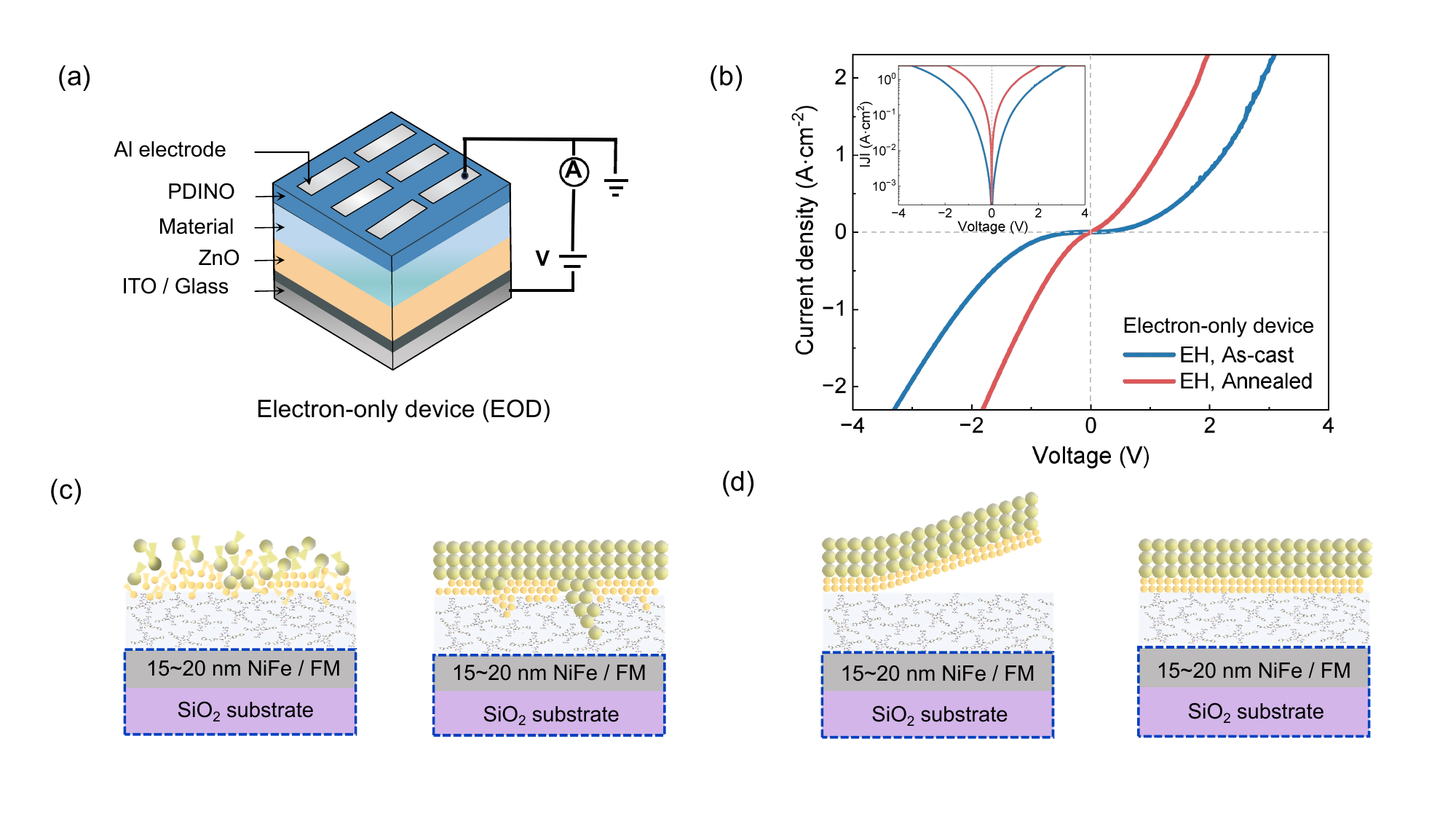}
      \caption{
  Vertical electron transport and top-electrode fabrication routes for EH-based devices.
(a) Schematic architecture and measurement configuration of the electron-only device (EOD) comprising ITO/glass, ZnO, T2-(NDI-T2)$_2$-2EH, PDINO, and Al layers.
(b) Room-temperature current density--voltage ($J$--$V$) characteristics of the as-cast and annealed EOD. The inset shows the corresponding semilogarithmic $|J|$--$V$ plots. Annealing was performed at 208\,°C for 10\,min.
(c,d) Schematic illustrations of OSV fabrication by (c) direct evaporation of the Co/Au top electrode onto the EH layer, illustrating possible metal penetration, and (d) transfer of a preformed Gr/Co/Au top electrode onto the EH layer.}
      \label{device_fabrication}
\end{figure}

Having established the annealing-induced structural changes in the T2-(NDI-T2)$_2$-2EH film, we first examined their implications for electrical properties using electron-only devices (EODs). The device architecture is shown in Figure~\ref{device_fabrication}a, and the room-temperature current density--voltage ($J$--$V$) characteristics are presented in Figure~\ref{device_fabrication}b. The nearly symmetric $J$--$V$ characteristics under forward and reverse bias indicate minimal rectification. Annealing increases the current density, indicating enhanced effective vertical electron transport. The effective vertical conductivity was determined from the low-bias differential conductance according to $\sigma=L(\mathrm{d}J/\mathrm{d}V)$, which follows from $J=\sigma V/L$ in the low-field regime, where $J$ is the current density, $V$ is the applied voltage, and $L$ is the organic-layer thickness. Using the differential slope at $V=0$ and an organic-layer thickness of 34~nm, the conductivity increases from $3.63 \times 10^{-8}$ to $1.32 \times 10^{-6}$\,S\,cm$^{-1}$ after annealing, corresponding to an approximately 36-fold enhancement (Table~1). Full details of the calculation are provided in the Supporting Information. Because this value is extracted from the complete ITO/ZnO/EH/PDINO/Al stack, it represents an effective device-level vertical conductivity rather than the intrinsic bulk conductivity of the EH layer. Nevertheless, the observed increase is consistent with the enhanced molecular order identified in Figure~\ref{structureandmorphology}~\cite{Coropceanu2007ChargeTransport}.

To examine how this annealing response translates to complete spin valves, we compared EH-based OSVs fabricated using two different top-electrode integration methods (Figure~\ref{device_fabrication}c,d). Direct evaporation exposes the EH surface to the metal flux and can produce a penetration-modified interphase (Figure~\ref{device_fabrication}c)~\cite{Guo2024Strain,Ding2018Reliable}. In contrast, the transfer method places a preformed Gr/Co/Au electrode onto EH without direct metal deposition (Figure~\ref{device_fabrication}d). This separation of electrode preparation from interface formation is also consistent with van der Waals and graphene-assisted transfer methods, in which preformed metal electrodes are transferred onto semiconductor surfaces to reduce damage caused by direct metal deposition~\cite{Liu2018VdWContact,Liu2022GrapheneTransfer}. The detailed transfer procedure is described in Section~S3 of the Supporting Information (SI). Magnetic measurements confirm that the Co/Au electrodes retained their hysteresis characteristics after transfer (Figure~S3). This approach avoids direct exposure of the EH surface to the evaporated metal flux, although the final contact still depends on how conformally the preformed electrode follows the organic surface.

\begin{figure}[h]
      \centering
      \includegraphics[width=\linewidth]{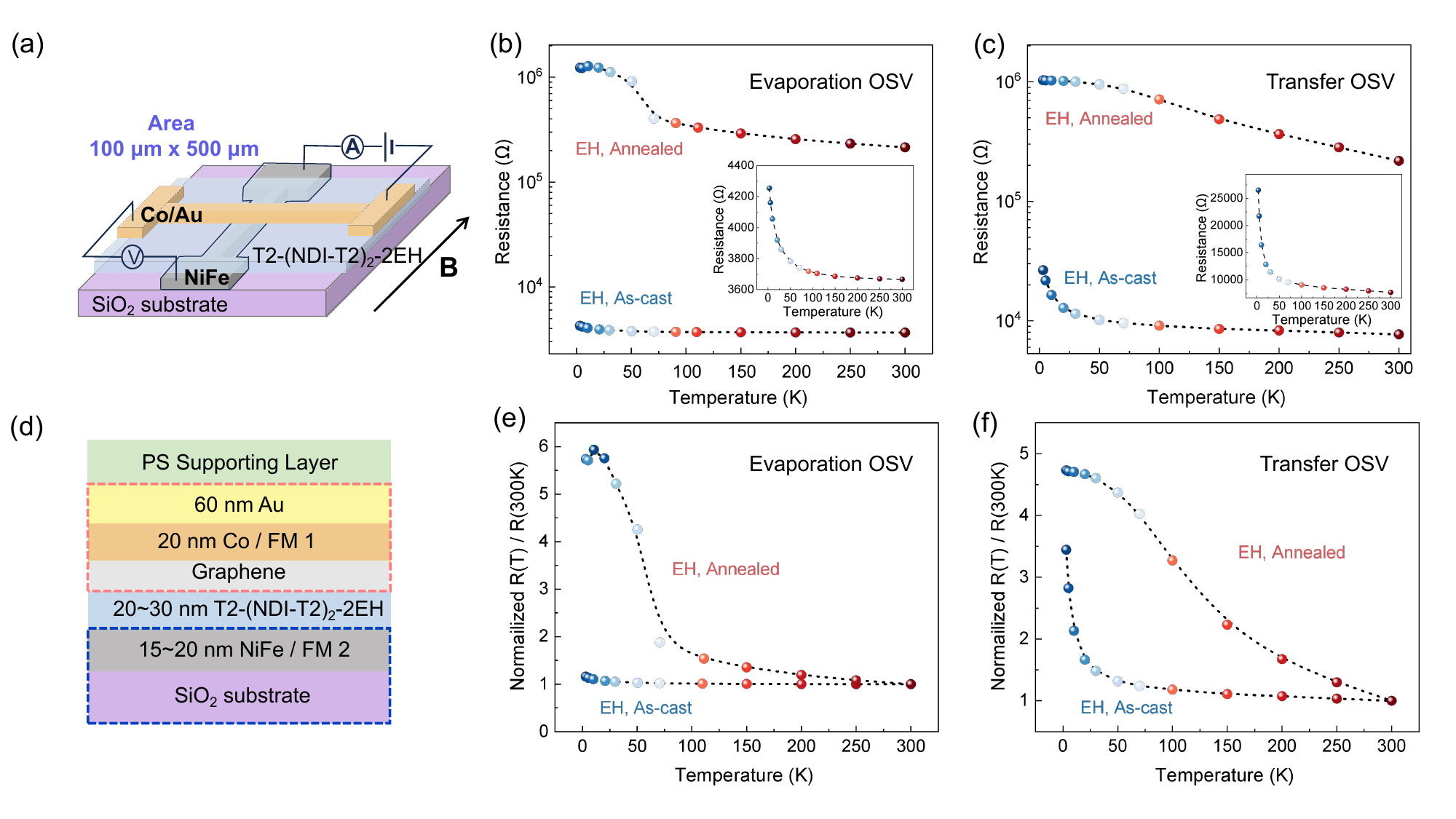}
      \caption{
Device configuration and temperature-dependent resistance of EH-based OSVs before and after annealing.
(a) Schematic of the OSV geometry and four-terminal resistance measurement configuration. The magnetic field was applied parallel to the device plane, while the current flowed perpendicular to the plane through an active area of 100\,$\mu$m $\times$ 500\,$\mu$m.
(b,c) Temperature-dependent resistance of OSVs fabricated by (b) direct top-electrode evaporation and (c) top-electrode transfer. The insets show enlarged views of the corresponding as-cast resistance data. (d) Layer structure of the OSV, comprising SiO$_2$/NiFe/EH/Gr/Co/Au, with the polystyrene (PS) supporting layer indicated. (e,f) Normalized $R(T)/R(300\,\mathrm{K})$ curves of OSVs fabricated by (e) direct top-electrode evaporation and (f) top-electrode transfer. The devices were first measured in the as-cast state and remeasured after in situ annealing at 208\,$^\circ$C for 10~min. The as-cast and annealed states are identified in each panel. Symbols represent the experimental data, while dotted lines are guides to the eye and do not represent fits.}
      \label{RT}
\end{figure}

Figure~\ref{RT}a shows the OSV geometry and resistance measurement configuration. The active area was 100\,$\mu$m $\times$ 500\,$\mu$m, and the magnetic field was applied parallel to the device plane during the field-dependent resistance measurements. Both architectures used the same NiFe bottom electrode, EH spacer, active area, and measurement configuration, with the top-contact architecture and its preparation method as the principal differences. Figure~\ref{RT}b,c compares the temperature-dependent resistances of the directly evaporated and transferred OSVs, respectively, before and after annealing. The layer structure of the transferred-electrode device is detailed in Figure~\ref{RT}d. To facilitate comparison of the temperature dependence, the corresponding resistance curves were normalized to their room-temperature values, $R(T)/R(300\,\mathrm{K})$, as shown in Figure~\ref{RT}e,f for the directly evaporated and transferred devices, respectively.

In contrast to the EOD response, the two completed OSVs become substantially more resistive after annealing. The as-cast directly evaporated device has a resistance of approximately $4.25~\mathrm{k\Omega}$ at 3~K and shows only a weak variation over the measured temperature range of 3--300\,K, as more clearly seen in the inset of Figure~\ref{RT}b. After annealing, its resistance reaches approximately $1230~\mathrm{k\Omega}$ at 3~K, corresponding to an approximately 290-fold increase (Table~1). The low resistance and weak temperature dependence of the as-cast directly evaporated device may reflect contributions from deposition-induced low-resistance pathways. Direct metal deposition can cause metal penetration or local damage within the organic layer, thereby reducing the effective transport distance and partially bypassing transport through the organic bulk~\cite{Vinzelberg2008TMR,Ding2018Reliable}. The transferred device exhibits a higher and more strongly temperature-dependent resistance in the as-cast state, increasing from approximately $7.5~\mathrm{k\Omega}$ at 300~K to $26.5~\mathrm{k\Omega}$ at 3~K, as shown more clearly in the inset of Figure~\ref{RT}c. After annealing, its resistance increases to approximately $220~\mathrm{k\Omega}$ at 300~K and $1040~\mathrm{k\Omega}$ at 3~K, corresponding to increases of approximately 29-fold and 39-fold, respectively. The normalized curves further show that annealing produces a stronger temperature dependence in both OSV architectures (Figure~\ref{RT}e,f). Thus, although annealing enhances the effective vertical conductivity of the EOD, it increases the resistance of both completed OSVs. Since the OSV resistance contains contributions from both the organic layer and the ferromagnet/organic spinterfaces, these opposing trends are consistent with increased contact or interfacial resistance after annealing. The common response of the two OSV architectures further indicates that this effect is not specific to the electrode-transfer process.

\begin{table}[htbp]
\captionsetup{justification=centering}
\caption{Annealing-induced changes in the room-temperature effective vertical conductivity of the EH electron-only device and the 3~K resistance of the two complete OSVs.}
\label{tbl:annealing}
\centering
\begin{tabular}{lccc}
\hline
Treatment &
\shortstack{Electron-only device: \\ conductivity, $\sigma$ ($\mathrm{S\,cm^{-1}}$)} &
\shortstack{Directly evaporated \\ OSV $R$ at 3~K ($\mathrm{k\Omega}$)} &
\shortstack{Transferred OSV \\ $R$ at 3~K ($\mathrm{k\Omega}$)} \\
\hline
As-cast & $3.63\times10^{-8}$ & 4.25 & 26.5 \\
Annealed & $1.32\times10^{-6}$ & 1230 & 1040 \\
Annealed/As-cast ratio & 36.5 & 290 & 39 \\
\hline
\end{tabular}
\end{table}

\begin{figure}[h]
      \centering
      \includegraphics[width=\linewidth]{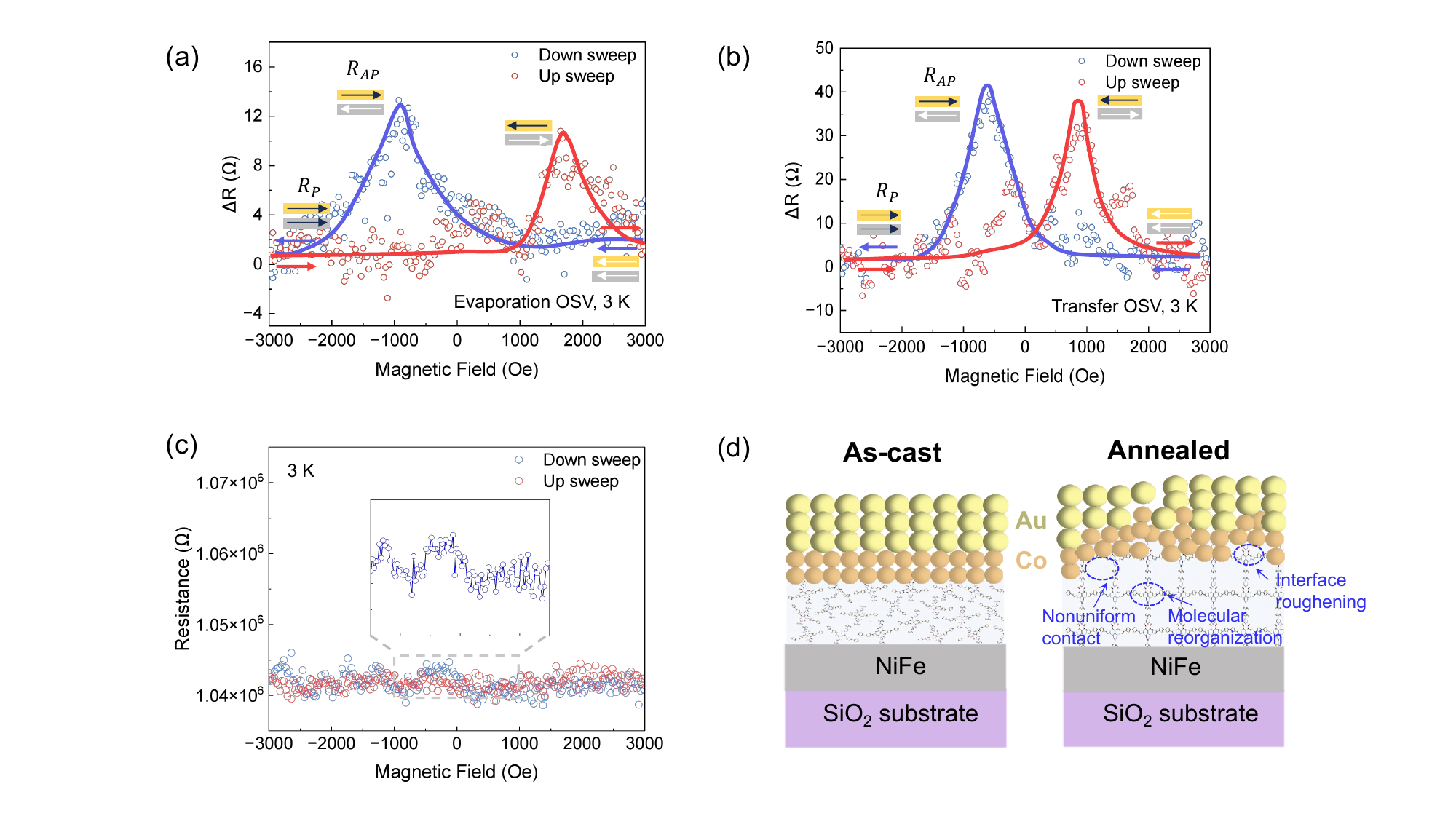}
      \caption{
Low-temperature magnetotransport of EH-based OSVs. $\Delta R$-$H$ curves of as-cast devices fabricated by (a) direct evaporation and (b) top-electrode transfer at 3~K. Blue and red symbols denote the down- and up-sweeps; solid lines are guides to the eye. (c) $R$-$H$ curves of the transferred OSV after annealing, measured at 3~K, showing no reproducible MR feature. (d) Possible annealing-induced molecular reorganization, interface roughening, and spatially nonuniform contact. The graphene layer is omitted for clarity.}
      \label{MR}
\end{figure}

The opposing annealing response is also pronounced in magnetotransport measurements. Before annealing, both OSV architectures exhibit reproducible sweep-dependent resistance peaks at 3~K (Figure~\ref{MR}a,b). At high magnetic fields, the magnetizations of the NiFe and Co electrodes are parallel, corresponding to the low-resistance state $R_{\mathrm{P}}$. Reversal of one electrode during the field sweep produces an antiparallel configuration and the high-resistance state $R_{\mathrm{AP}}$~\cite{Xiong2004,Ke2026Temperature}. The field-dependent resistance change is defined as $\Delta R(H) = R(H) - R_{\mathrm{sat}}$, where $R_{\mathrm{sat}}$ is the reference resistance determined from the high-field saturation regions and is used as an experimental approximation to the parallel-state resistance $R_{\mathrm{P}}$. The MR ratio is calculated as $\Delta R(H)/R_{\mathrm{sat}} \times 100\%$, with its peak value corresponding to the resistance difference between the high- and low-resistance magnetic configurations. The directly evaporated device shows broader peaks with $\Delta R$ of approximately 10--13~$\Omega$, whereas the transferred device shows sharper peaks with $\Delta R$ of approximately 35--40~$\Omega$. The corresponding fractional MR values are approximately 0.2--0.3\% and 0.13--0.15\%, respectively, because the transferred device has a higher baseline resistance. The corresponding raw resistance, resistance change, and MR curves obtained from the same measurement are compared in Figure~S6. The larger absolute \(\Delta R\) and more clearly resolved features of the transferred device are consistent with a more spatially defined top contact after eliminating direct metal deposition~\cite{Guo2024Strain,Ding2018Reliable}. The down- and up-sweeps show resistance peaks at negative and positive magnetic fields, respectively, characteristic of a hysteretic spin-valve response commonly observed in OSV systems~\cite{Wagemans2011SpinTransport}. As the temperature increases from 3 to 70~K, the positive- and negative-field MR peaks progressively shift toward zero field, consistent with the temperature-dependent narrowing of the electrode coercive fields and with characteristic OSV magnetoresistance behavior~\cite{Ke2026Temperature}. The MR peaks remain experimentally resolvable up to 70~K; above this temperature, no reproducible MR feature can be distinguished from the background fluctuations within the measurement sensitivity. The complete temperature-dependent measurements are provided in the Supporting Information.

However, after in situ annealing, no reproducible MR peaks could be resolved for either the directly evaporated or transferred EH OSVs under the measurement conditions. Figure~\ref{MR}c shows the annealed transferred device at 3~K, for which only irregular resistance fluctuations remain around a baseline of approximately 1040~k$\Omega$. The temperature-dependent MR curves of the directly evaporated device before annealing are provided in Figure S7.

Because each device was measured before and after annealing under the same current and temperature, the change does not originate from a comparison between separately fabricated samples. It also cannot be explained simply by the loss of ferromagnetic switching. The transferred Co/Au electrode retains magnetic hysteresis after transfer, and the EH/Co/Au stack retains magnetic switching before and after annealing at both 300 and 3~K (Figures~S3 and S5). The absence of reproducible MR therefore coincides with the large increase in electrical resistance and is consistent with annealing-induced changes in spin injection, transport, or detection within the completed device.

We therefore propose that the contrasting responses of the EOD and the complete OSVs originate from the different effects of annealing on bulk molecular order and the spinterfaces, as illustrated in Figure~\ref{MR}(d). The relatively smooth surface of the as-cast film provides comparatively uniform contact conditions. After annealing, the films exhibit increased roughness and domain formation, as shown in Figure~\ref{structureandmorphology}(c). If comparable morphological reorganization occurs at the OSV interfaces, it could result in spatial nonuniformity, reduce the effective contact area, and redistribute local current pathways~\cite{Willenbockel2015Interface}. In the directly evaporated device, annealing may additionally affect the penetration-modified spinterface~\cite{Zhan2008Energy,Ding2018Reliable}. These architecture-dependent microscopic changes can produce the same macroscopic outcome: increased complete-device resistance despite improved effective vertical transport through organic material.

Spin injection and detection also depend on spin-dependent hybridization and effective spin polarization at the spinterfaces~\cite{Barraud2010Unravelling,Cinchetti2017Activating,Raman2013Interface,Luo2023MolecularPacking,Droghetti2016Dynamic,Steil2013Spinterface}. Changes in molecular packing and contact nonuniformity can therefore modify the interface spin polarization, while increased roughness can enhance interface-related spin scattering. More broadly, molecular-level control of the local electronic structure can strongly tune spin-dependent charge transport, as recently demonstrated by metal-center engineering in molecular junctions
~\cite{Ke2026CCSChem}. Although the spinterfaces were not imaged directly, the common increase in resistance and absence of reproducible MR after annealing in both architectures show that improved bulk order alone is insufficient to predict complete-device performance. Electrode transfer improves the definition of the initial MR response, but it does not eliminate the device's susceptibility to subsequent morphological reorganization.

In summary, we compared directly evaporated and transferred-electrode OSVs based on the same NDI-conjugated oligomer. Both as-cast architectures exhibit MR features, while the transferred device provides sharper peaks and a larger absolute \(\Delta R\). Annealing enhances the structural order and effective vertical conductivity of the material but increases the resistance of both completed OSVs and renders their MR features absent. These findings show that the magnetotransport properties of an OSV depend not only on bulk molecular order but also on the formation and thermal stability of the spinterfaces. Optimization must therefore improve the organic semiconductor while maintaining contact continuity, uniform current pathways, and stable interface spin polarization. Future work should focus on thermally robust and conformal spinterface engineering that preserves contact homogeneity during molecular reorganization, thereby translating improved bulk transport into stable spin-dependent device performance.

\section{Experimental}

\subsection{Materials and Oligomer Synthesis}
The discrete n-type oligomer T2--(NDI--T2)$_2$--2EH (C$_{84}$H$_{86}$N$_4$O$_8$S$_6$; calculated molar mass: 1472.0~g~mol$^{-1}$) was synthesized by adapting the palladium-catalyzed direct-arylation procedure reported in Ref.~\cite{Matsidik2023Evolution}. The product was purified by column chromatography and characterized by $^1$H NMR spectroscopy. All other materials and solvents were used as received unless otherwise stated.

\subsection{Differential Scanning Calorimetry}
The thermal behavior of T2-(NDI-T2)$_2$-2EH was examined using a DSC 2500 calorimeter (TA Instruments) under nitrogen, following the previously reported measurement procedure~\cite{Matsidik2023Evolution}. Samples of approximately 2--5~mg were subjected to two heating--cooling cycles at a rate of $10~\mathrm{K\,min^{-1}}$ in both directions. Melting parameters were extracted from the second heating scan to reduce the influence of the initial thermal history, while crystallization parameters were obtained from the first cooling scan. The corresponding thermograms and thermal transition parameters are presented in the Supporting Information.

\subsection{Film Preparation and Structural Characterization}
Companion films of T2--(NDI--T2)$_2$--2EH for atomic force microscopy (AFM) and X-ray reflectivity (XRR) measurements were spin-coated at 3000~rpm onto Si substrates bearing a 300~nm thermally grown SiO$_2$ layer, the thickness of organic-layer is approximately 30~nm. These independently prepared, electrode-free films were used to characterize the morphology and structural organization of the organic semiconductor in the as-cast and annealed states. Annealing was performed at 208~$^\circ$C for 10~min, using the same annealing temperature and duration as for the completed OSVs. AFM topography was acquired under ambient conditions using a SmartSPM 1000 system (AIST-NT) equipped with an NSG10 cantilever with a nominal tip radius of 10~nm. XRR measurements were performed using a Rigaku SmartLab diffractometer equipped with a 9~kW rotating Cu anode and Cu K$\alpha$ radiation in a parallel-beam configuration. A K$\beta$ filter was used during the measurements, and the reflectivity curves were fitted using GenX 3, version 3.6.27~\cite{Glavic2022GenX3}.

\subsection{Electron-Only Device Fabrication and Electrical Characterization}
Electron-only devices with the architecture ITO/ZnO (approximately 30~nm)/T2--(NDI--T2)$_2$--2EH (approximately 34~nm)/PDINO (approximately 20~nm)/Al (approximately 100~nm) and an active area of 0.04~cm$^2$ were used to evaluate vertical electron transport. The completed
devices were annealed in situ at 208~$^{\circ}$C for 10~min. Room-temperature current density--voltage ($J$--$V$) characteristics were measured under nitrogen using a Keithley 236 SourceMeter. The effective vertical conductivity was extracted from the low-bias differential conductance according to $\sigma=L(\mathrm{d}J/\mathrm{d}V)$ at $V=0$, using an organic-layer thickness $L$ of 34~nm.

\subsection{Organic Spin Valve Fabrication and Magnetotransport Measurements}
Two vertical OSV architectures were investigated: NiFe/EH/Co/Au with a directly evaporated top electrode and NiFe/EH/Gr/Co/Au with a transferred top electrode. Both architectures had an EH spacer thickness of approximately 20--30~nm and an active junction area of 100~$\mu$m $\times$ 500~$\mu$m. The bare Si/SiO$_2$ substrates were treated with oxygen plasma for 10~min prior to deposition of the NiFe bottom electrodes. T2--(NDI--T2)$_2$--2EH was dissolved in chloroform (99.9\% purity) at a concentration of 5.6~mmol\,L$^{-1}$, and the solution was spin-coated onto the NiFe bottom electrodes at 1000~rpm. For the directly evaporated devices, the Co/Au top electrode was deposited directly onto the EH layer. For the transferred-electrode devices, a preformed Gr/Co/Au top electrode was prepared and laminated onto the EH layer following the previously reported graphene-assisted electrode-transfer procedure~\cite{Ke2026Temperature}. Further details of the transfer process are provided in Supporting Information Section~S3. For each architecture, the same individual device was characterized before and after in situ annealing at 208~$^\circ$C for 10~min. Temperature-dependent resistance measurements were performed from 3 to 300~K. Four-terminal magnetotransport measurements were carried out using a Physical Property Measurement System (PPMS; Quantum Design), with the magnetic field applied parallel to the device plane and the current flowing perpendicular to the plane. Magnetic-field-dependent resistance curves were recorded at selected temperatures between 3 and 70~K using constant currents of 0.015, 0.020, and 0.025~$\mu$A.

\section*{Author Contributions}
Z.L. and Y.K. contributed equally to this work. Z.L.: Investigation, data curation, formal analysis, and writing for the original draft. Y.K.: Investigation, data curation, and formal analysis. H.K.: Investigation and resources. S.G.: Investigation and writing---review \& editing. R.E.: Investigation, formal analysis, and visualization. O.H.: Supervision and resources. F.O.: Writing---review \& editing. M.S.: Supervision and resources. G.S.: Conceptualization, supervision, resources, and writing---review \& editing. S.D.: Conceptualization, supervision, resources, and writing---review \& editing.
\section*{Conflict of interest}
The authors declare no conflict of interest.
\section*{Acknowledgements}
The authors acknowledge funding by the Deutsche Forschungsgemeinschaft (DFG, German Research Foundation) under TRR 386, project B5 (Project No. 514664767), the National Key R\&D Program (2024YFA1209600), the National Natural Science Foundation of China (52373250), and the Fundamental and Interdisciplinary Disciplines Breakthrough Plan of the Ministry of Education of China (JYB2025XDXM410).
\section*{Supporting information}

Differential scanning calorimetry thermograms and thermal parameters; magnetic characterization of the NiFe bottom electrode, Co/Au electrodes before and after transfer, and EH/Co/Au stacks before and after annealing; top-electrode transfer procedure; electron-only device fabrication, energy-level diagram, and effective vertical conductivity determination; and additional current- and temperature-dependent magnetoresistance measurements of devices with directly evaporated and transferred top electrodes.

The following files are available free of charge.

%%%%%%%%%%%%%%%%%%%%%%%%%%%%%%%%%%%%%%%%%%%%%%%%%%%%%%%%%%%%%%%%%%%%%
%% If you are using classical BibTeX rather than biblatex,
%% remove the \printbibliography and uncomment the \bibliograpy one
%%%%%%%%%%%%%%%%%%%%%%%%%%%%%%%%%%%%%%%%%%%%%%%%%%%%%%%%%%%%%%%%%%%%%

\printbibliography
%\bibliography{acs-template.bib}

\end{document}